\documentclass[unnumsec,webpdf,contemporary,large]{oup-authoring-template}%
\graphicspath{{Fig/}}

\usepackage{siunitx}
\usepackage{multirow}

\usepackage{color}
\usepackage{hyperref, soul}
\definecolor{linkColor}{rgb}{1,0,0}
\definecolor{citeColor}{rgb}{1,0,0}
\hypersetup{pdfborder={0 0 0}, colorlinks=true,urlcolor=linkColor,citecolor=citeColor}

\begin{document}

\journaltitle{Microscopy and Microanalysis}
\DOI{DOI HERE}
\copyrightyear{2023}
\pubyear{2023}
\access{Advance Access Publication Date: Day Month Year}
\appnotes{Paper}

\firstpage{1}

%\subtitle{Subject Section}

\title[Probing defectivity in 2D hBN using TEM-EELS]{Probing defectivity beneath the hydrocarbon blanket in 2D hBN using TEM-EELS}

\author[1,2,3,$\ast$]{Dana O. Byrne\ORCID{0000-0000-0000-0000}}
\author[3]{Jim Ciston\ORCID{0000-0000-0000-0000}}
\author[2,3,4,$\ast$]{Frances I. Allen\ORCID{0000-0000-0000-0000}}

\authormark{Byrne et al.}

\address[1]{\orgdiv{Department of Chemistry}, \orgname{University of California, Berkeley}, \orgaddress{\state{CA}, \postcode{94720}, \country{USA}}}
\address[2]{\orgdiv{Department of Materials Science and Engineering}, \orgname{University of California, Berkeley}, \orgaddress{\state{CA}, \postcode{94720}, \country{USA}}}
\address[3]{\orgdiv{National Center for Electron Microscopy, Molecular Foundry}, \orgname{Lawrence Berkeley National Laboratory}, \orgaddress{\state{CA}, \postcode{94720}, \country{USA}}}
\address[4]{\orgdiv{California Institute for Quantitative Biosciences}, \orgname{University of California, Berkeley}, \orgaddress{\state{CA}, \postcode{94720}, \country{USA}}}

\corresp[$\ast$]{Corresponding authors. \href{email:email-id.com}{dana$\_$byrne@berkeley.edu, francesallen@berkeley.edu}}

\received{Date}{0}{Year}
\revised{Date}{0}{Year}
\accepted{Date}{0}{Year}

%\editor{Associate Editor: Name}

%\abstract{
%\textbf{Motivation:} .\\
%\textbf{Results:} .\\
%\textbf{Availability:} .\\
%\textbf{Contact:} \href{name@email.com}{name@email.com}\\
%\textbf{Supplementary information:} Supplementary data are available at \textit{Journal Name}
%online.}

\abstract{
%\boxedtext{
The controlled creation and manipulation of defects in 2D materials has become increasingly popular as a means to design and tune new material functionalities. However, defect characterization by direct atomic imaging is often severely limited by surface contamination due to a blanket of hydrocarbons. Thus, analysis techniques are needed that can characterize atomic scale defects despite the contamination. In this work we use electron energy loss spectroscopy to probe beneath the hydrocarbon blanket, characterizing defect structures in 2D hexagonal boron nitride (hBN) based on fine structure in the boron K-edge. Since this technique is performed in a transmission electron microscope, imaging can also be used to assess contamination levels and other factors such as tears in the fragile 2D sheets, which can affect the spectroscopic analysis. Furthermore, by locally probing individual areas, multiple regions on the same specimen that have undergone different defect engineering treatments can be investigated for systematic studies at increased throughput. For 2D hBN samples irradiated with different ions for a range doses, we find spectral signatures indicative of boron--oxygen bonding that can be used as a measure of sample defectiveness depending on the ion beam treatment. 
%In addition, two different sample cleaning methods are employed in order to investigate their effects on the local chemical structure of the defects as well as the effect of the hydrocarbons themselves on the boron K-edge signals.
}
\keywords{2D materials, defects, EELS, contamination}

%In this paper, we demonstrate TEM-EELS as a method to characterize defects in ion irradiated monolayer and few-layer hBN.

% \boxedtext{
% \begin{itemize}
% \item Key boxed text here.
% \item Key boxed text here.
% \item Key boxed text here.
% \end{itemize}}

\maketitle
\section{Introduction}

Defect engineering of 2D materials is used to tune material properties and to impart new functionalities for a range of emerging applications, including in energy storage~\citep{Liu2020}, catalysis~\citep{Sun2019}, quantum technology~\citep{Bertoldo2022}, and for the desalination of water~\citep{Safaei2020}. For example, substitutional doping of 2D hexagonal boron nitride can be used to create dopant-vacancy complexes with unique single-photon emission behavior~\citep{Mendelson2021}, and through the generation of clusters of atom vacancies (nanopores) in a 2D material, selective filters for atoms, ions, and small molecules can be realized~\citep{Liu2021,Fang2019}. Various techniques for introducing defects into the 2D crystal lattice have been developed, including chemical and thermal treatments, exposure to plasma, dielectric breakdown, and electron/ion irradiation~\citep{Su2021,Wang2017,Jiang2019a}. Across the board, robust methods for characterizing the resulting atom-scale defects are needed. In the case of graphene, Raman spectroscopy has proven to be an invaluable characterization tool allowing high-throughput screening with a high level of sensitivity to defect type due to the rich structural information contained within peak intensity ratios ~\citep{Eckmann2012ProbingSpectroscopy,Beams2015RamanGraphene}. However, for many other 2D materials and their defects, distinct Raman spectral signatures do not exist and one must turn to alternative characterization methods.  

Direct imaging of 2D materials at atomic resolution can be achieved using (scanning) transmission electron microscopy ((S)TEM)~\citep{Meyer2008,Krivanek2010} and scanning probe microscopy (SPM)~\citep{Zhang2008}. In addition to generating atom maps, (S)TEM spectroscopy can be used to determine the chemical identity of individual atoms, and various advanced (S)TEM/SPM techniques can be used to extract information on electronic, vibrational, optical, magnetic and mechanical properties~\citep{NgomeOkello2021,Musumeci2017AdvancedMaterials,Ophus2019}.
A key prerequisite for high-resolution imaging of the 2D lattice is that the sample area should not be obscured by contaminants. However, surface contamination with a blanket of hydrocarbons is common, resulting from the transfer process (e.g.\ polymer residue) and also simply from exposure of the samples to air. Indeed, so-called adventitious carbon forms on samples in air in a matter of minutes~\citep{Barr1995}. Multi-layer samples can typically still be imaged by (S)TEM at atomic resolution because the atomic columns boost transmission signal, but this is generally not the case for weakly scattering monolayer samples. 
Various recipes and approaches have been reported to reduce the amount of hydrocarbon contamination on 2D material samples~\citep{Dean2010BoronElectronics,Garcia2012EffectiveDevices,Dyck2018MitigatingStudies,Zabelotsky2023ExploringCrystal}. However, these techniques are often sample-specific, variably effective, and may require specialized equipment such as lasers in ultrahigh vacuum~\citep{Tripathi2017}. In short, the researcher is often still faced with the task of searching for clean patches of sample before imaging can begin. During imaging, the total area obscured by hydrocarbon contamination can also increase even further, for example, due to electron-beam-induced deposition by the focused STEM probe~\citep{Dyck2018MitigatingStudies}. For imaging on atomic length scales, the size of clean regions can be small (a few square nanometers is sufficient to image tens of atoms). Yet for representative surveys of larger areas, the hydrocarbon blanket is still a major limiting factor.

A technique is thus needed that allows one to probe the defectivity of 2D materials in a high throughput manner by ``seeing through'' the surface contamination in order to obtain statistically relevant information on local bonding configurations. In the work presented here, we propose spectroscopic characterization by spatially-integrated TEM-based electron energy-loss spectroscopy (EELS) to meet this need. 

EELS is analogous to X-ray absorption spectroscopy (XAS), in that both techniques probe core electron excitations into unoccupied orbitals and can thus be used to evaluate local electronic structure~\citep{Goode2015}. Synchrotron-based monochromated X-ray sources have been available for longer than monochromated electron sources; therefore the spectral resolution of XAS has historically been far superior to that of EELS (tens of meV vs.\ hundeds of meV). Correspondingly, the use of XAS to characterize local bonding configurations is more established. However, the development of new monochromated electron sources for (S)TEM has brought EELS into the running, with highly specialized electron microscopes enabling a spectral resolution down to the meV level~\citep{Krivanek2019}. In terms of spatial resolution, soft X-ray beamlines for scanning transmission X-ray microscopy (STXM) typically deliver probe sizes of a few tens of nanometers, and by implementing the diffraction-based ptychography approach, a spatial resolution down to a few nanometers has been achieved~\citep{Shapiro2014}. Even so, in XAS, atomic scale fingerprints are by default averaged, since the effective probe size is still relatively large. In contrast, EELS mapping is typically performed in STEM mode, where the probe can be focused to the sub-{\AA}ngstrom length scale. This allows elemental and chemical mapping on a single atom level. However, as is the case for STEM imaging, atomic resolution STEM-EELS of 2D monolayers demands clean samples, so that the spectroscopic signal from the individual atoms of interest is not swamped by noise. 

In our TEM-EELS approach, the working principle is to broaden the beam to collect signal from large numbers of defects in one acquisition and hence probe beneath the blanket (as in XAS), but at the same time, to make use of concurrent TEM imaging to select suitable regions to analyze. In this way, the presence of e.g.\ larger defects that may develop in the fragile 2D materials can be factored into the analysis of the spectroscopy data and independent regions of the same specimen that have undergone different defect engineering treatments can be probed individually, for efficient systematic surveys of sample defectiveness. 

The 2D material we investigate in this paper is hexagonal boron nitride (hBN) and we employ an ion irradiation method to introduce the atomic scale defects that we then probe using TEM-EELS. The irradiation is performed site-selectively using a focused ion beam (FIB) microscope, rastering the beam at low dose to create vacancy defects (ejected atoms) due to single-ion hits~\citep{Thiruraman2018}. We use a multibeam FIB that can deliver helium, neon or gallium ions for this step. Lighter ions have lower sputter yields due to their lower momentum transfer, meaning they induce less damage to the material, while heavier ions have larger sputter yields and deal more damage. Thus by modulating both ion type and irradiation dose, sample defectiveness can be controlled. In terms of the spectroscopy, the boron K-edge is known to be rich in chemical information and XAS studies conducted elsewhere have revealed distinct spectral signatures in the boron K-edge of 2D hBN that have been attributed to particular defect structures~\citep{Huber2015,McDougall2017}. Our TEM-EELS surveys also reveal these spectral peaks, the relative intensities of which depend on the ion beam treatment parameters. We demonstrate that TEM-EELS can be used to probe these variations despite the hydrocarbon contamination on the specimen,
%\cite{Egerton2011ElectronMicroscope}. 
which means that the TEM-EELS method can be used for defect characterization when direct imaging of the defects is not a viable option. Since the throughput of TEM-EELS is much higher than that of atomic resolution mapping, the method can also be used to screen the effect of different defect engineering treatments in order to narrow down on parameters. Subsequently, STEM-EELS at atomic resolution can be performed (assuming samples are sufficiently clean), in order to directly correlate the spectral signals with their defect structures.

\section{Materials and methods}

\subsection{Preparation of few-layer and monolayer hBN}

Custom holey silicon nitride (SiNx) TEM substrates were fabricated in-house from \SI{3}{mm} silicon chips with \SI{20}{nm}-thick free-standing SiNx membranes (window size \SI{20}{\micro\m}) obtained from Norcada Inc. Using a custom pattern, arrays of \SI{\sim200}{nm} diameter apertures together with index marks were drilled into each SiNx membrane using a \SI{50}{\pA}, \SI{30}{keV} focused Ga ion beam (Zeiss ORION NanoFab He-Ne-Ga FIB microscope). The hBN samples were transferred directly onto these membrane supports as described below. 

Few-layer hBN samples were obtained from pristine bulk hBN crystals using the standard tape exfoliation method. These flakes were then transferred to the custom holey SiNx TEM substrates using mechanical dry release polydimethylsiloxane (PDMS) gel transfer~\citep{Castellanos-Gomez2014}. The flake thicknesses as measured by atomic force microscopy (AFM) were approximately 3.33--6.66\,nm, corresponding to 10--20 layers. 

Monolayer hBN samples were prepared via electrochemical delamination transfer of chemical vapor deposition (CVD) hBN monolayers grown on copper foil obtained from Grolltex Inc. In the electrochemical delamination method, monolayer hBN on copper is spin-coated with polymethyl methacrylate (PMMA) and clipped into a two-electrode system with NaCl electrolyte and a glassy carbon cathode. A \SI{5}{V} bias generates hydrogen bubbles at the anode, delaminating the hBN/PMMA film from the copper \citep{Wang2011}. The hBN/PMMA films were then rinsed with water before being transferred to the custom holey SiNx TEM substrates and cleaned with acetone to remove the PMMA.

\subsection{Ion irradiation to create defects}

In order to generate a set of samples with varying levels of defectivity, we used He, Ne and Ga ions from the Zeiss ORION NanoFab multibeam FIB operated in ion `showering' mode (a low-dose raster). A range of defect densities in the monolayer and few-layer hBN samples were generated using a range of target doses: 50, 100, and 500 ions/nm$^2$ for \SI{25}{keV} He ions, 1, 10, 20, and 50 ions/nm$^2$ for \SI{25}{keV} Ne ions, and 0.1 and 1 ions/nm$^2$ for \SI{25}{keV} Ga ions. Scan parameters of \SI{1}{\micro\s} dwell time, variable scan spacing, and variable numbers of repeat scans were used to enable fine control over the spatial range and position of the irradiations, allowing for a single sample to be irradiated with multiple ions and doses. 

\subsection{TEM and TEM-EELS measurements}

The TEM images and EELS data were collected using a double-aberration-corrected modified FEI Titan 80-300 microscope (the TEAM I instrument at the Molecular Foundry, LBNL) equipped with a high-resolution Gatan Imaging Filter Continuum K3 direct electron detector spectrometer system. The microscope was operated at \SI{80}{kV} accelerating voltage. With source monochromation, a \SI{0.21}{eV} spectral resolution was obtained. High-resolution TEM imaging conditions used an approximate dose rate of 300--600\,e$^-$/\AA$^2$/s and EELS measurements were taken at approximate dose rates of 300--400\,e$^-$/\AA$^2$/s with total accumulated doses ranging from 15,000--80,000\,e$^-$/\AA$^2$. The diameter of the illuminated area for the TEM-EELS acquisitions was set to \SI{\sim200}{nm}.

\subsection{Data analysis}

Analysis of the EELS data was conducted using the open source Python library HyperSpy~\citep{HyperSpy}. Background signal contributions were removed using power law background subtraction and all B K-edge spectra were normalized to the B onset peak at \SI{192}{eV}. Finally, all spectra were smoothed using total variation filtering with the L2 term minimization weight ranging from 0.1 to 5, for the high and low signal-to-noise spectra, respectively.

\section{Results and discussion}

\subsection{Pristine vs.\ Ne-irradiated few-layer hBN}

Hydrocarbon surface contamination on freshly transferred hBN samples formed islands up to \qty{100}{nm} in size, leaving similarly sized patches of uncontaminated hBN. This level of contamination is typical for 2D material samples before any post-transfer cleaning treatments have been applied and is attributed to various sources, including polymers from the transfer process and hydrocarbons in the environment \citep{Yang2022EffectMaterials,Peng2017InfluenceMaterials}. 
 
\begin{figure*}
     \centering
     \includegraphics[width=0.8\textwidth]{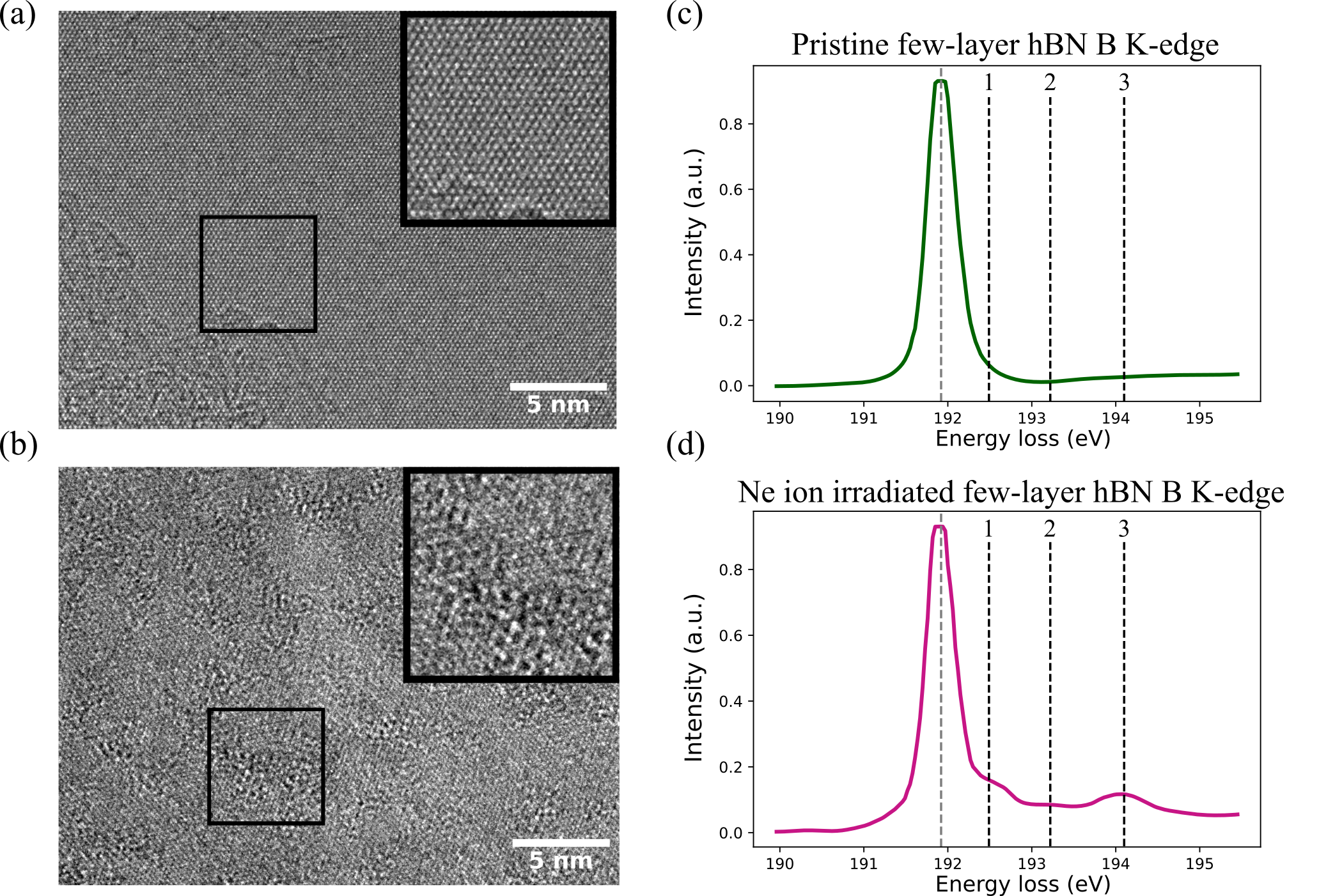}
     \caption{TEM images of (a) pristine and (b) Ne-irradiated few-layer hBN, with high magnification insets. Monochromated TEM-EELS B K-edge results for the (c) pristine and (d) Ne-irradiated samples (corresponding to the regions imaged in (a) and (b), respectively). The vertical dashed lines indicate the experimentally determined positions of peaks 1, 2, and 3. The Ne irradiation dose was \SI{50}{ions/nm^2}.}
  \label{fig:Few_Layer_EELS_insets}
\end{figure*}

Figure~\ref{fig:Few_Layer_EELS_insets}(a) shows a representative TEM image of an as-transferred pristine few-layer hBN sample, showing surface contamination and contamination-free regions. The high-magnification inset shows a contamination-free area; here, the pristine hBN lattice is clearly visible. For samples that had been irradiated with ions, contamination coverage increased significantly, severely hampering direct imaging of the ion-irradiated hBN lattice. This is shown for an Ne-ion-irradiated sample in Fig.~\ref{fig:Few_Layer_EELS_insets}(b). Beam-induced hydrocarbon contamination under scanning focused ion (and electron) probes is a well-known phenomenon that is caused by the mobilization and decomposition of residual hydrocarbons already present on the sample and in the sample chamber~\citep{Hlawacek2013}. This form of contamination is difficult to avoid unless working with ultraclean samples in ultrahigh vacuum.

The extensive hydrocarbon blankets on the ion-irradiated samples almost entirely obscured the underlying hBN, thus making TEM imaging of the ion-induced damage to the hBN lattice extremely challenging. Instead we turn to EELS for spectroscopic characterization of the defects created by the ion shower. When performing EELS of these samples in STEM mode, the contamination over the scanned region invariably gets worse, since the focused probe locally fixes mobile surface contaminants in place. In contrast, the lower dose of broad-beam illumination in TEM mode did not have this effect. Thus in our investigations we performed EELS in TEM mode, allowing the collection of spectra over broader areas (to increase signal-to-noise ratios) without adding additional contamination to the region of interest. Unlike XAS, TEM-based EELS allows direct imaging of the specimen at high resolution in the same experiment, enabling the user to precisely select the regions for analysis. Therefore, multiple regions on a single specimen that have undergone different ion beam treatments can be surveyed individually. Furthermore, the TEM images can be used to assess the condition of the sample in terms of any tears or larger vacancy defect clusters that may be present and could affect the resulting spectra.

Figures~\ref{fig:Few_Layer_EELS_insets}(c) and (d) show monochromated B K-edge TEM-EELS results corresponding to the pristine few-layer and Ne-ion-irradiated few-layer hBN samples, respectively. These spectra were collected for the same regions shown in the TEM images of Figs.~\ref{fig:Few_Layer_EELS_insets}(a) and (b). Despite the obscuring contamination in the ion-irradiated case, the TEM-EELS analysis shows a stark change in spectral signature after ion irradiation, with the emergence of three additional peaks in the B K-edge spectrum. This spectral change can be attributed to a change in the local boron environment due to the introduction of defects, as described further below.  

Theoretical and experimental work performed elsewhere (using X-ray absorption near-edge spectroscopy (XANES)) indicates that fine structure peaks near the B K-edge of hBN are likely due to point defects involving substitutional atoms on nitrogen sites~\citep{Caretti2011,Huber2015,McDougall2017,Niibe2010}. These fine structure peaks are also referred to as chemical shifts and are caused by local changes in electronegativity and bonding around the atom species being probed. Generally it is understood that fine structure peaks for hBN appearing after the onset of the B K-edge involve various configurations of substitutional oxygen doping. In contrast, fine structure peaks before the onset of the B K-edge are generally attributed to carbon doping.
%These peak shifts can essentially be traced back to the element electronegativities relative to nitrogen; oxygen is more electronegative, whereas carbon is less electronegative. 
In our spectra, the fine structure peaks labeled ``1", ``2" and ``3" in Fig.~\ref{fig:Few_Layer_EELS_insets}(d) occur after the onset of the B K-edge and agree well with the literature values for B-ON$_2$, B-O$_2$N, and B-O$_3$, respectively (see Table~\ref{table:1}). Thus our results indicate that in our hBN samples B-O bonding is present whereas B-C bonding is not, despite the hydrocarbon blanket. 

Thus from this first set of EELS data we can deduce that Ne ion irradiation at a dose of \SI{50}{ions/nm^2} caused notable damage to the few-layer hBN sample and resulted in the formation of substitutional oxygen impurity defects. In the following, the relative intensities of the BN$_3$ main peak and the B-O$_{x}$N$_{3-x}$ ($x=1,2,3$) defect peaks will be used to compare defectivity and bonding in few-layer and monolayer hBN samples that have undergone various ion irradiation treatments.

\subsection{Few-layer hBN irradiated with He, Ne, and Ga ions} 

\begin{figure}
     \centering
     \includegraphics[width=\linewidth]{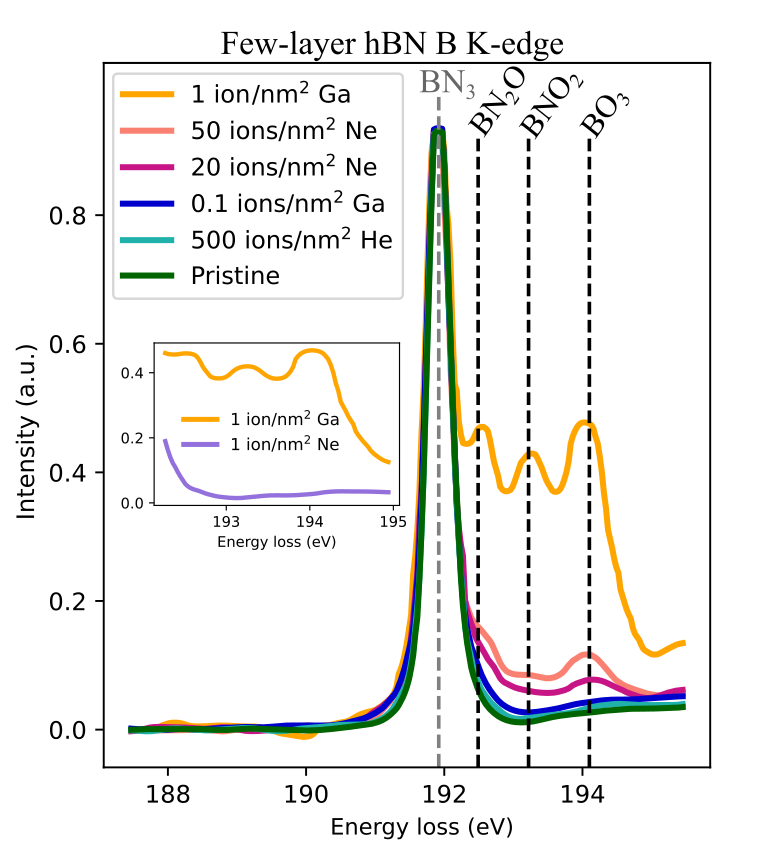}
     \caption{B K-edges for pristine, He, Ne, and Ga irradiated few-layer hBN flakes. Dashed vertical lines indicate the experimentally determined positions of the defect peaks and their assigned boron environments. The inset shows a zoomed-in direct comparison of the fine structure for a sample exposed to Ga and Ne ion showers to the same dose of \SI{1}{ion/nm^2}.}
  \label{fig:Few_layer_Ne_vs_Ga}
\end{figure}

In Fig.~\ref{fig:Few_layer_Ne_vs_Ga}, the monochromated B K-edge spectra for different regions of a few-layer hBN sample that had been irradiated with He, Ne or Ga ions using a range of doses are compared. Since sputter yield scales with ion mass, Ga ions were showered at the lowest doses, with Ne and He ions showered at increasingly higher doses. Also included in the figure are the spectra for the pristine and \SI{50}{ions/nm^2} Ne-ion-irradiated samples from Fig.~\ref{fig:Few_Layer_EELS_insets}. Each spectrum is normalized to the BN$_3$ peak (at \SI{192}{eV}) so that the relative intensities of the respective defect peaks can be compared.

The most intense defect peaks are observed for hBN that had been irradiated using Ga ions to a dose of \SI{1}{ion/nm^2}, with each defect peak reaching approximately half of the intensity of the primary BN$_3$ peak. We thus infer that this Ga ion beam treatment lead to a significant amount of B--O bonding in the material. The inset of Fig.~\ref{fig:Few_layer_Ne_vs_Ga} compares this Ga ion result with that of a sample irradiated using Ne ions at the same dose. As expected, the intensity of the defect peaks (i.e.\ the degree of defectivity) is higher after irradiation with the heavier Ga ions. However, the extent of the observed increase is much greater than collision cascade calculations predict. For example, calculating using SRIM code~\citep{Ziegler2010} for 20-layer hBN, we obtain vacancy generation yields of \SI{15}{vacancies/ion} for \SI{25}{keV} Ne ions and \SI{55}{vacancies/ion} for \SI{25}{keV} Ga ions; i.e.\ the yield increases by a factor of 3.7. (Impacted atoms will move to interstitial sites while a few will be forward sputtered and thus removed from the material.) In contrast, the intensities of the defect peaks for Ne vs.\ Ga ions shown in the figure inset each increase by a factor of over 10. This discrepancy may be due to the relative instability of the Ga-irradiated sample under the electron beam, which was observed for few-layer samples irradiated with Ga ions to doses of \SI{1}{ion/nm^2} and above. For example, hBN few-layer flakes irradiated with Ga ions to \SI{5}{ion/nm^2} rapidly formed large holes and etched away even under relatively low electron dose rates (50\,e$^-$/\AA$^2$/s). This Ga-induced instability means that such samples will be more susceptible to structural and chemical changes post ion irradiation. In these cases, the TEM-EELS approach needs to be applied with caution, since it may not be gentle enough to avoid damaging highly defective samples. However, we note that defective hBN samples can also be unstable in air~\citep{Dai2023}, hence even if they are probed completely non-invasively, the structures may no longer be representative of those formed directly after irradiation. 
%Hydrocarbon surface contamination has also been found to accelerate etching and therefore may contribute to the degradation of Ga irradiated the sample. 

Returning to the main part of Fig.~\ref{fig:Few_layer_Ne_vs_Ga}, hBN flakes irradiated with Ne ions to doses of \SI{50}{ions/nm^2} and \SI{20}{ions/nm^2} have the next highest peak intensities after the intense \SI{1}{ion/nm^2} Ga result. Next come the spectra for Ga ion irradiation at \SI{0.1}{ions/nm^2} and He ion irradiation at \SI{500}{ions/nm^2}. The spectrum for the latter only deviates slightly from that of the pristine hBN flake, indicating that very little defectivity was introduced by the light ions despite the relatively high dose. Indeed, for 20-layer hBN and \SI{25}{keV} He ions, SRIM predicts a vacancy generation yield of just \SI{0.6}{vacancies/ion}~\citep{Ziegler2010}.

%Monolayer hBN flakes did, however have a small increase in peak 1 signal after He ion irradiation (Fig. \ref{fig:Monolayer_Ion_irr_comparison}).

\subsection{Monolayer hBN irradiated with He, Ne and Ga ions}

\begin{figure}
     \centering
     \includegraphics[width=\linewidth]{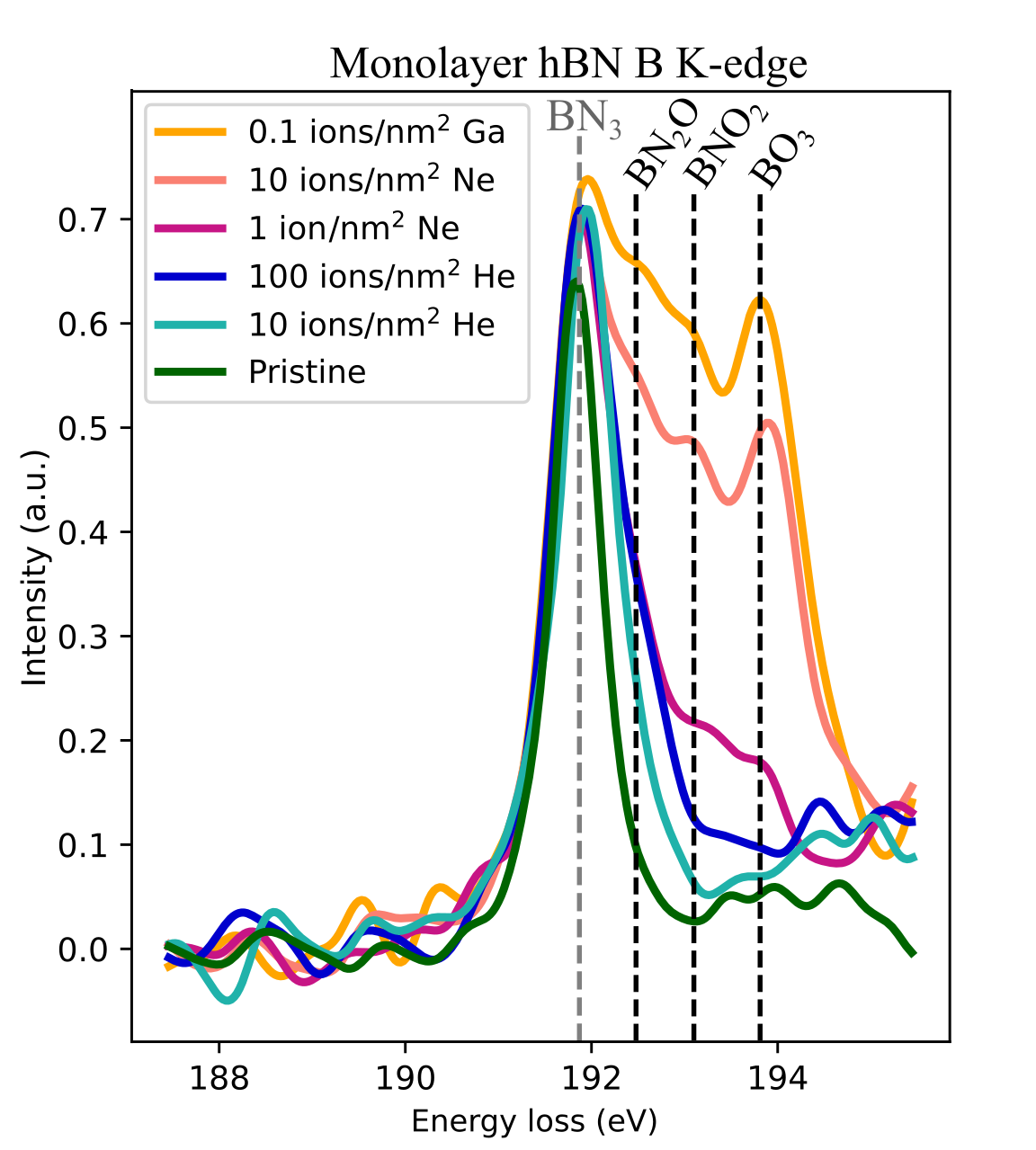}
     \caption{B K-edges for pristine, He, Ne, and Ga irradiated monolayer hBN samples. Dashed vertical lines indicate the experimentally determined positions of the defect peaks and their assigned boron environments.}
  \label{fig:Monolayer_Ion_irr_comparison}
\end{figure}

The ion shower doses used for the monolayer hBN samples were overall lower than for the few-layer samples in order to preserve the integrity of the more fragile monolayers. It was found that Ne and Ga ion irradiation at doses above \SI{10}{ions/nm^2} and \SI{0.5}{ions/nm^2}, respectively, tended to damage the suspended monolayer to such a degree that the film would immediately break under the electron beam and become impossible to image. The ion doses were therefore kept below the sample destruction limit. Prolonged electron exposure of the ion-irradiated monolayers was found to ultimately break suspended regions even for minimally defective monolayer samples, so extra caution was taken to limit the total electron dose used for these TEM-EELS acquisitions.

Fig.~\ref{fig:Monolayer_Ion_irr_comparison} shows the monochromated B K-edge TEM-EELS results for monolayer hBN irradiated using a range of He, Ne, and Ga ion doses. All spectra are once again normalized to the BN$_3$ peak (smoothing was performed after normalization, causing the slight variation in BN$_3$ peak intensities). The spectrum for pristine monolayer hBN is also included. The signal-to-noise ratio of the monolayer spectra is lower than for the few-layer samples, which is expected, because lower electron doses were used for these measurements. Furthermore, the monolayer is by nature much thinner and therefore there are fewer inelastic scattering events that produce spectral features. Despite this, the B K-edge fine structure peaks indicated by the dashed vertical lines reveal distinct differences in defectivity for the ion-irradiated monolayer samples. 

Similarly to the few-layer hBN results, the most intense fine structure features are observed for a Ga-ion-irradiated sample (here for a dose of \SI{0.1}{ions/nm^2}). The intensities of the defect peaks for the highest dose Ne-ion-irradiated sample (\SI{10}{ions/nm^2}) are close behind. We again attribute these enhanced defect peaks to the greater instability of these samples, both of which were observed to contain large pores (of the order of \SI{10}{nm^2}), which readily expanded and ultimately lead to fracture of the membranes after prolonged imaging. 

Comparing directly with the B K-edge results for the few-layer samples in Fig.~\ref{fig:Few_layer_Ne_vs_Ga}, the defect peaks in Fig.~\ref{fig:Monolayer_Ion_irr_comparison} are proportionally more intense, despite the lower ion doses used for the irradiation. This indicates that the ion shower treatments ultimately caused more damage to the monolayer samples. This is interesting, because the probability for vacancy generation decreases as the sample becomes thinner, as will now be discussed. 

For a single-layer sample, all vacancy-producing collisions will result in sputtering of the impacted atoms from the material, therefore we can look to the literature and compare sputter yields computed using atomistic simulations of keV ion irradiation of monolayers. For example, for a monolayer graphene target, the sputter yields for \SI{25}{keV} He and Ne ions have been computed to be around \SI{5e-4}{atoms/ion} and \SI{0.16}{atoms/ion}, respectively~\citep{Lehtinen2010,Yoon2016b}. By extrapolation, the sputter yield for \SI{25}{keV} Ga ions would be around \SI{0.5}{atoms/ion}. These values are two to three orders lower than the vacancy generation yields discussed previously for the few-layer hBN samples. We thus conclude that the defectivity probed by the TEM-EELS shown in Fig.~\ref{fig:Monolayer_Ion_irr_comparison} is enhanced due to a combination of the inherent instability of defective hBN monolayers in air~\citep{Dai2023} and beam-induced sputtering in the TEM~\citep{Meyer2009} despite the low dose rates used. In both cases, the defective monolayers are likely to be more unstable than the respective few-layers, resulting in their larger measured defectivity. Nevertheless, apart from the \SI{0.1}{ions/nm^2} Ga-irradiated monolayer, the trend in defect peak intensities essentially follows sputter yield and dose considerations. While TEM-EELS acquisition can increase the defect peaks, the ability to directly image the samples means that any changes induced by the electron beam can be factored into the analysis with careful and consistent control of the electron dose. Moreover, any regions of the specimen that have become severely damaged (e.g.\ upon exposure to air before they enter the TEM) can immediately be identified.

Table~\ref{table:1} lists the peak positions (in eV) obtained from the B K-edge spectra for the ion-irradiated few-layer and monolayer hBN samples and compares with literature values obtained by XANES of a multi-layer pyrolytic hBN sample~\citep{Caretti2011}. Peaks 1, 2 and 3 correspond to the various BO$_x$N$_y$ bonding configurations, as discussed previously, while peak 0 corresponds to BN$_3$. Our peak energies and standard error values are determined by averaging the measured peak positions for all five of the ion-irradiated few-layer spectra from Fig.~\ref{fig:Few_layer_Ne_vs_Ga} and for three of the monolayer spectra (those with the most intense defect peaks) from Fig.~\ref{fig:Monolayer_Ion_irr_comparison}. Peak fitting was performed using a Gaussian model in HyperSpy with automatic background recognition. Good agreement with the literature peak positions is found, especially in the case of the the few-layer. For the monolayer, there is indication of a shift to lower energies for the higher order defect peaks. Although the shifts are small, they may be evidence of subtle changes in local electronic structure dependent on the dimensionality of the hBN (number of layers), which may affect the structural configuration of the defects. When surveying literature values, one would therefore need to be careful to differentiate data obtained from few-layer vs.\ monolayer samples.
\begin{table}
\centering
\caption{B K-edge fine structure peak energy positions for various BO$_x$N$_y$ bonding configurations, comparing the TEM-EELS values obtained for ion-irradiated few-layer and monolayer hBN samples in the present work with literature values from experimental XANES studies of pyrolytic hBN.}
\label{table:1}
\begin{tabular}{m{2em} m{3.2em} m{5.5em} m{5.9em} m{5.9em}} 
     \hline \hline
     \multirow{2}{2em}{Peak label} &\multirow{2}{3.2em}{Boron en\-vi\-ron\-ment} & \multirow{2}{5.5em}{Peak energy values from XANES (eV)\footnotemark[1] }& \multicolumn{2}{c}{Peak energies from TEM-EELS} \\ \cline{4-5}
      &  &   & Few-layer hBN (eV) & Monolayer hBN (eV) \\ [5.5ex]  
     \hline
      0 & B-N$_3$ & $191.98 \pm 0.02$ & $191.92 \pm 0.01$ & 191.87 $\pm$ 0.03 \\ 
      1 & B-ON$_2$ & $192.62 \pm 0.02$ & 192.49 $\pm$ 0.07 & 192.48 $\pm$ 0.06 \\
      2 & B-O$_2$N & $193.26 \pm 0.02$ & 193.22 $\pm$ 0.05 & 193.10 $\pm$ 0.04 \\
      3 & B-O$_3$ & $193.9 \pm 0.02$ & 194.1 $\pm$ 0.05 & 193.81 $\pm$ 0.10 \\
     \hline\hline
     \multicolumn{5}{l}{\footnotemark[1]\citep{Caretti2011}}
\end{tabular}
\end{table}

\subsection{Testing methods to remove the hydrocarbon blanket}

The ultimate goal is to image the various hBN defect structures directly and in order to do this, the surface hydrocarbon layer needs to be removed (preferably, it should not present in the first place). A key point to consider here is whether contamination removal treatments also alter the local chemical environments of the hBN defects. Therefore in this final section we have tested two cleaning treatments and use TEM-EELS of the B K-edge to track any changes in the defect peaks as a result of these treatments. 

One method used to locally remove surface hydrocarbon contamination from a sample inside the TEM is to shower electrons over selected areas in a broad-beam illumination condition~\citep{Li2021}. Similar electron beam shower methods are in fact more commonly used to locally deposit hydrocarbons to form a barrier to surface diffusion of mobile contamination~\citep{Egerton2004}. However, for the conditions implemented here (electron dose rate of 1000 e/\AA$^2$/s and accumulated dose of up to 1.2$\times$10$^6$\,e$^-$/\AA$^2$), visual inspection by TEM imaging before and after the electron beam shower treatment shows that the area of contamination coverage was significantly reduced (see Fig.~S1). Furthermore, TEM-EELS analysis of the C K-edge confirms that carbon was removed (Fig.~S2). The cleaning process presumably involves inelastic interactions of the electrons with the organic structures, resulting in bond breaking~\citep{Egerton2004} and subsequent removal of the hydrocarbon molecule fragments into the TEM vacuum.
%Hettler2017,Mitchell2015

\begin{figure*}
     \centering
     \includegraphics[width=\textwidth]{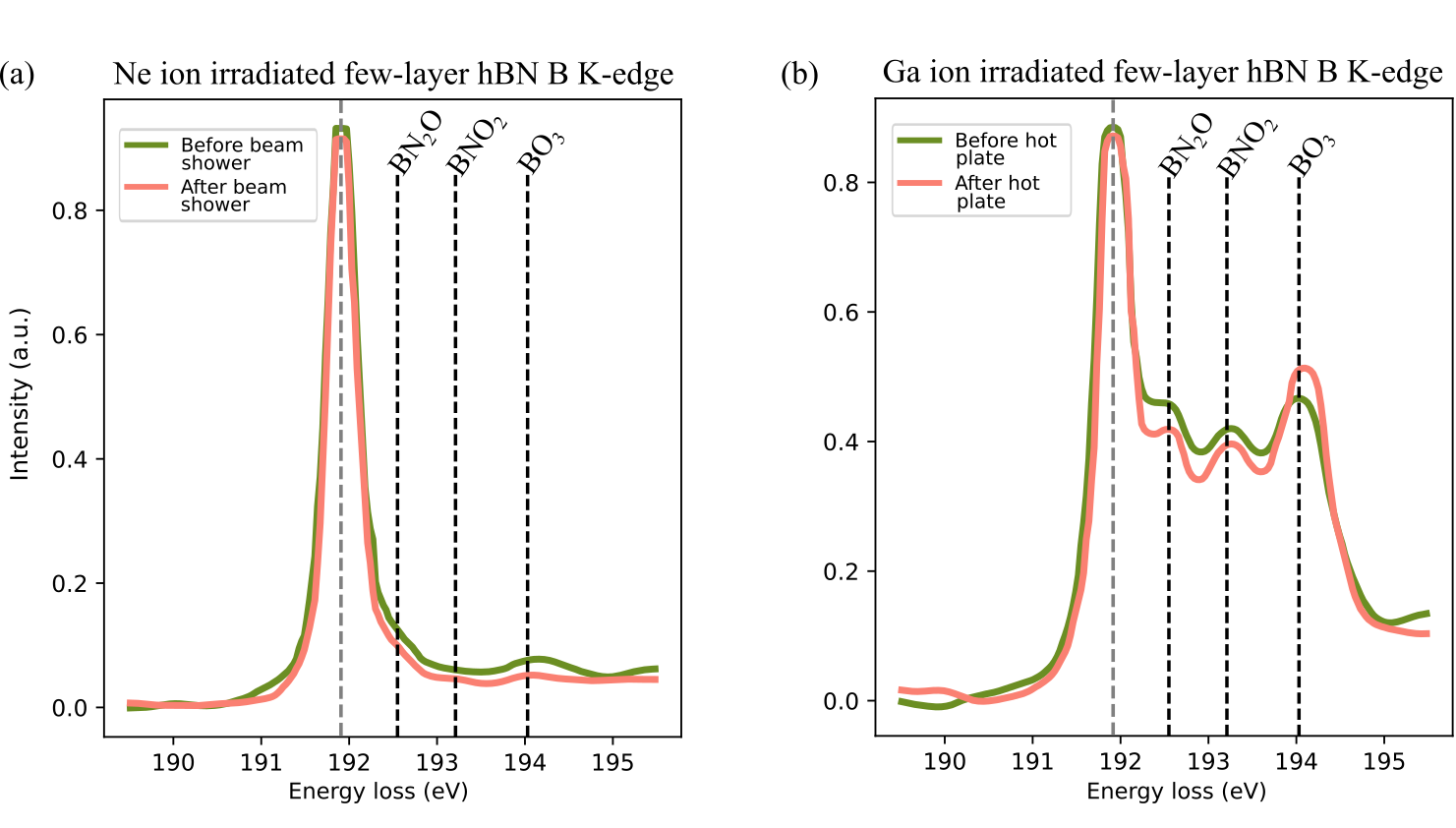}
     \caption{B K-edges for (a) Ne ion irradiated (\SI{20}{ions/nm^2}) few-layer hBN before and after electron beam shower and (b) Ga ion irradiated (\SI{1}{ion/nm^2}) few-layer hBN before and after hotplate treatment in air.}
  \label{fig:Cleaning_EELS}
\end{figure*}

In Fig. \ref{fig:Cleaning_EELS}(a), monochromated TEM-EELS results for the B K-edge before and after the electron beam shower are compared. The sample in this case was few-layer hBN that had been irradiated with Ne ions to a dose of \SI{20}{ions/nm^2}. We note that the ion beam treatment used here was relatively mild, because few-layer hBN samples that had been exposed to more severe ion beam treatments were easily destroyed during the electron beam shower step due to electron beam-induced etching of the unstable edge structures. Therefore, while the electron beam shower method is effective at removing hydrocarbon contamination, it may not be ideal for preserving the initial defectivity of the underlying hBN sample, especially in the case of more defective samples. For the Ne-irradiated sample investigated here, we see that after the electron beam shower, the intensities of all three defect peaks only decrease slightly. This points to removal of bonding O impurities from the hBN during the cleaning treatment, although the relative fractions of BN$_2$O, BNO$_2$, and BO$_3$ configurations appear to stay roughly the same.  

Another cleaning treatment that is sometimes used to remove hydrocarbon contamination from 2D material samples is heating the samples in an oxygen environment~\citep{Garcia2012EffectiveDevices}. In our experiments we took inspiration from this method and heated \SI{1}{ion/nm^2} Ga-irradiated few-layer hBN on a hot plate at \SI{400}{\celsius} in air for \SI{15}{mins}. After this treatment, we observed that the contamination islands reduced in size and that the intensity of the C K-edge decreased significantly, confirming the effectiveness of the hotplate method for contamination removal.

In Fig~{\ref{fig:Cleaning_EELS}}(b), the B K-edge monochromated EELS results from the hotplate cleaning test are compared. We see that the hotplate treatment lowers the intensities of peaks 1 and 2, indicating a reduction in BN$_2$O and BNO$_2$ bonding configurations. However, peak 3 increases in intensity. A possible reason for the relative increase of peak 3 is the formation of B$_2$O$_3$ precipitate, a complex with almost the same B K-edge transition energy as BO$_3$, and which has been theorized to contribute to the intensity of peak 3 observed in XANES studies~\citep{Huber2015, Caretti2011}. In contrast, the relative intensity of peak 3 does not seem to be affected by the electron beam shower (as seen in in Fig.~\ref{fig:Cleaning_EELS}(a)), which may be due to the reduced levels of oxygen in vacuum and/or electron beam irradiation inhibiting reconfiguration. 

In summary, neither cleaning method preserved the initial defect peaks observed in the B K-edge spectra of the ion irradiated samples. This result underscores the need for spectroscopic techniques like TEM-EELS that can probe beneath the contamination, and can thereby be used to optimize cleaning treatments so that the initial chemical structures of interest are maintained. 

\section{Conclusions}

Hydrocarbon contamination of 2D materials is an ongoing challenge for defect characterization by direct imaging. However, in this work we show that hydrocarbon-obscured atomic-scale defects in 2D hBN can be probed spectroscopically using monochromated TEM-EELS B K-edge analysis. Using He, Ne and Ga ion irradiation at controlled dose, variably defective few-layer and monolayer hBN samples were fabricated that were then probed using the TEM-EELS approach. This analysis revealed defect peaks corresponding to substitutional doping of nitrogen atoms with oxygen. The defect peak intensities scaled with increasing defectivity. The TEM-EELS method thus allowed classification of defects and a comparison of defectivity depending on the ion beam treatment. Both the few-layer and monolayer hBN samples showed similar trends after ion beam treatment, indicating that dimensionality on these length scales does not greatly affect the defect chemistry. 

By combining TEM-EELS with low-magnification (and hence low-dose) TEM imaging in the same experiment, a more comprehensive understanding of the sample structure and its relationship to the resulting spectra can be obtained. For example, particularly fragile hBN samples from irradiation with Ga ions could be identified, with the observed larger holes explaining the particularly intense defect peaks measured by TEM-EELS in those cases. While such samples tended to be unstable under electron beam illumination, we speculate that structural and chemical changes can also occur before analysis, which needs to be considered regardless of characterization approach.

In order to achieve atomic resolution imaging of the 2D lattice and its defects, one can attempt to apply one or more of the various methods reported in the literature for removing the blanketing hydrocarbon contamination. We tested two of these methods and obtained direct spectroscopic evidence showing that the defect chemistry can be altered during these treatments. Further studies need to be conducted in this area in order to optimize cleaning treatments, which is where TEM-EELS analysis can play a central role. Ideally, of course, samples will be free from surface contamination from the outset. However, even in such cases we propose that TEM-EELS can be useful for screening the defectivity of samples. Subsequently, smaller regions can be probed at atomic resolution using STEM-EELS in order to directly correlate individual defect structures with their spectroscopic signals.  

%where did O come from in the first place? do we need to speculate?

\section{Competing interests}

No competing interest is declared.

%\section{Author contributions statement}

%Must include all authors, identified by initials, for example:
%S.R. and D.A. conceived the experiment(s),  S.R. conducted the experiment(s), S.R. and D.A. analysed the results.  S.R. and D.A. wrote and reviewed the manuscript.

\section{Acknowledgments}
This work was funded in part by NSF Award No.\ 2110924. D.O.B.\ also acknowledges funding from the Department of Defence though the National Defence Science \& Engineering Graduate (NDSEG) Fellowship Program. Work at the Molecular Foundry was supported by the Office of Science,
Office of Basic Energy Sciences, of the U.S. Department of Energy under Contract No.\ DE-AC02-05CH11231. J.C.\ acknowledges additional support from the Presidential Early Career Award for Scientists and Engineers (PECASE) through the U.S. Department of Energy.

\bibliographystyle{abbrvnat}
\bibliography{refs_new}
%\bibliography{references,refs_new}

%https://www.overleaf.com/project/6424b2b079736d2bcee88b48

%USE THE BELOW OPTIONS IN CASE YOU NEED AUTHOR YEAR FORMAT.
%\bibliographystyle{abbrvnat}
%\bibliography{reference}

\end{document}

\begin{frontmatter}

\title{\texorpdfstring{Supplementary Material:\\ \vspace{2mm} Probing defectivity beneath the hydrocarbon blanket in 2D hBN using TEM-EELS}}

\author[label1,label2,label3]{Dana O. Byrne}

\address[label1]{Department of Chemistry, UC Berkeley, Berkeley, CA 94720, USA}
\address[label2]{Department of Materials Science and Engineering, UC Berkeley, Berkeley, CA 94720, USA}
\address[label3]{National Center for Electron Microscopy, Molecular Foundry, LBNL, Berkeley, CA 94720, USA}
\address[label4]{California Institute for Quantitative Biosciences, UC Berkeley, Berkeley, CA 94720, USA}

\author[label3]{Jim Ciston}

\author[label2,label3,label4]{Frances I. Allen}

\end{frontmatter}

\begin{figure}[h]
    \centering
    \includegraphics[width=\linewidth]{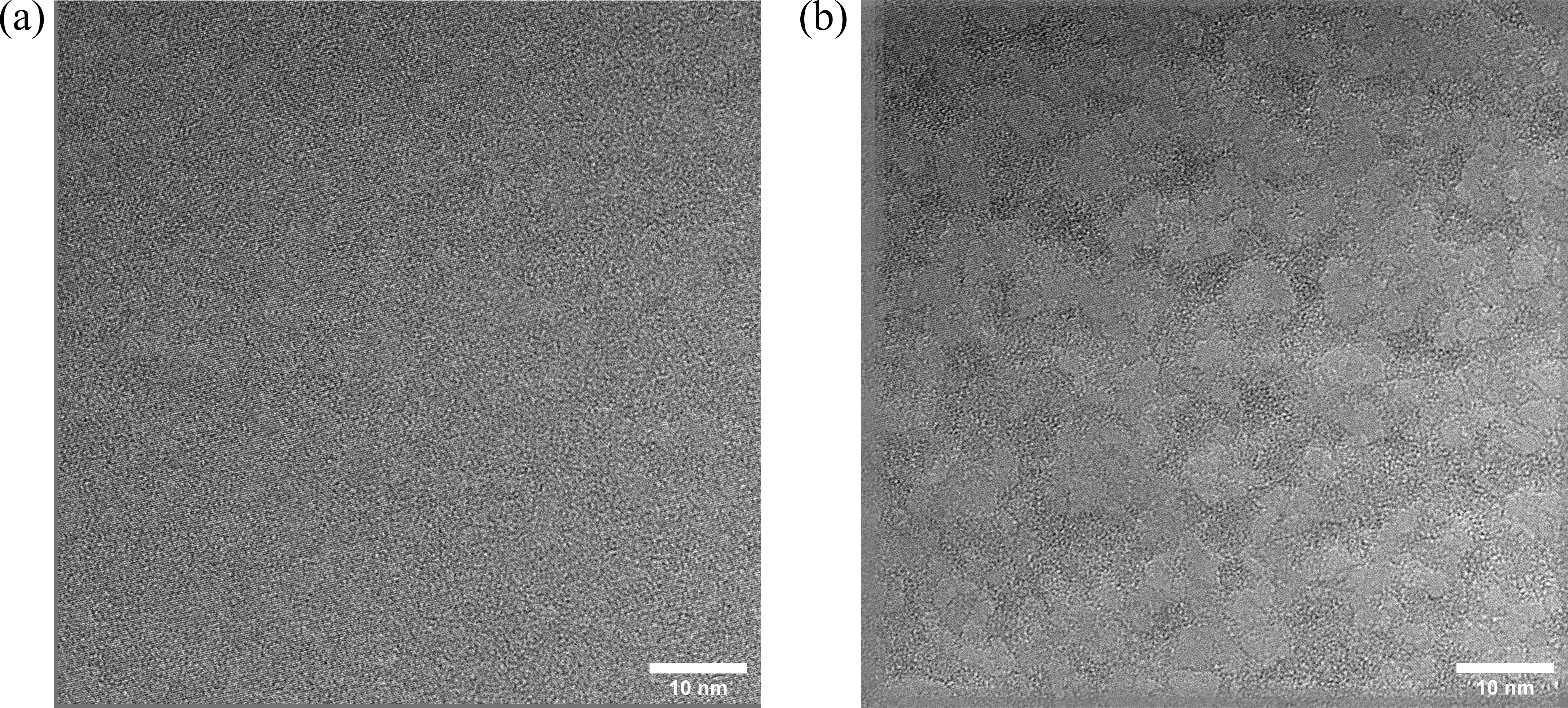}
    \caption{TEM images of Ne irradiated few-layer hBN (a) before and (b) after electron beam shower cleaning treatment. Beam shower cleaning was carried out with a dose of 400 \,e$^-$/\AA$^2$/s for 20 min. The Ne irradiation dose was \SI{20}{ions/nm^2}.}
    \label{}
\end{figure}

\begin{figure}[h]
    \centering
    \includegraphics[width=\linewidth]{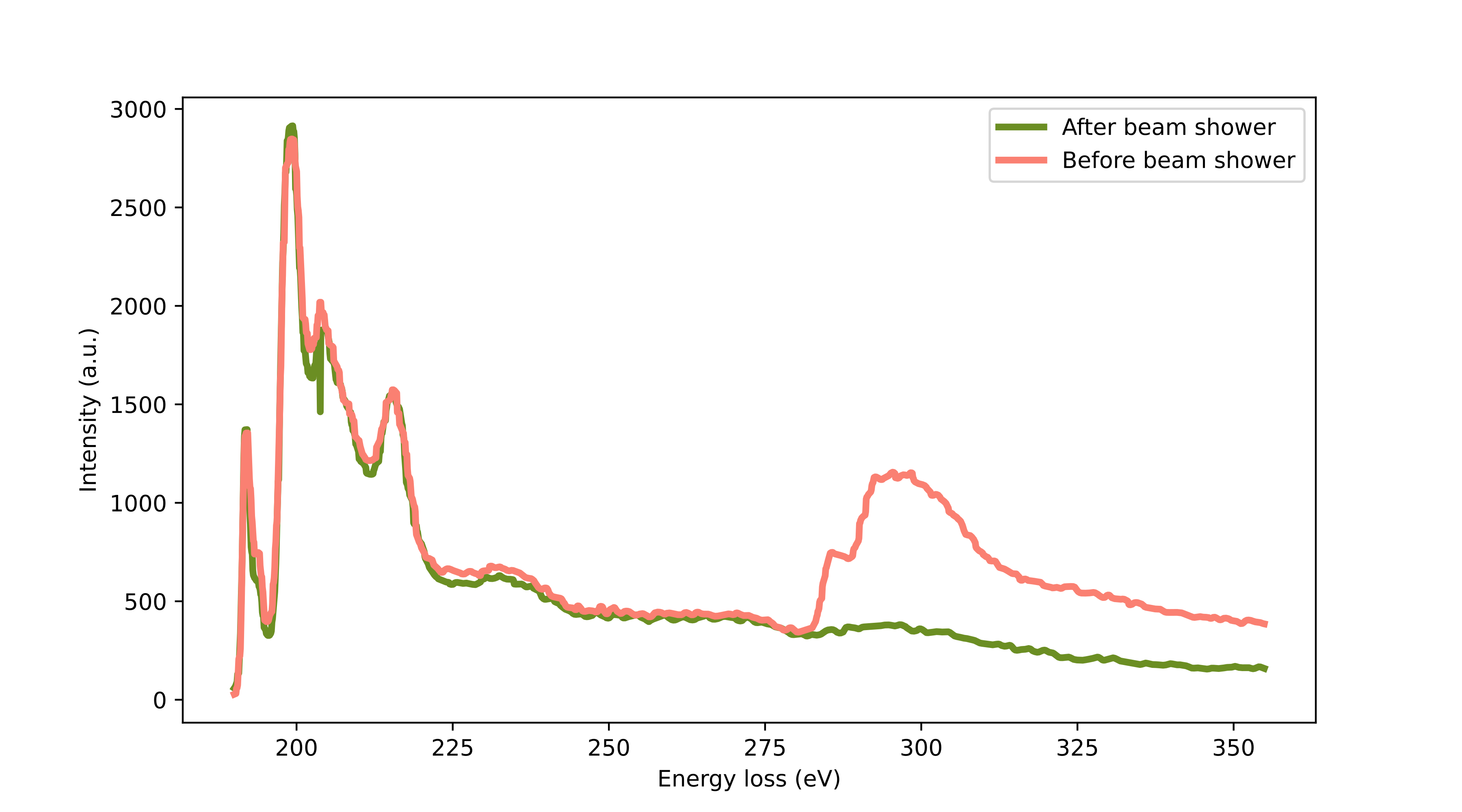}
    \caption{EEL spectra of the B and C K-edges of the same Ne irradiated sample shown in Fig.\ S1 before and after electron beam shower cleaning. All C K-edge spectra underwent background subtraction, smoothing with 200 weight, and normalization to the B K-edge onset peak.}
    \label{}
\end{figure}